\documentclass{elsarticle}
\pdfoutput=1
\usepackage{hyperref}
\usepackage{txfonts}
\usepackage{graphicx}
\usepackage{natbib}
\usepackage{epsf}

\usepackage{numcompress}

\begin{document}
\begin{frontmatter}

\title{ CoRoT pictures transiting exoplanets \\}

\author{ Claire Moutou$^{1,2}$ and Magali Deleuil$^2$}

\address{$^1$ Canada-France-Hawaii Telescope Corporation, 65-1238 Mamalahoa Highway, Kamuela HI 96743, moutou@cfht.hawaii.edu \\
$^2$ Laboratoire d'Astrophysique de Marseille, 38 rue Fr\'ed\'eric Joliot-Curie, 13376 Marseille cedex 13, France, magali.deleuil@lam.fr}

\begin{abstract}
The detection and characterization of exoplanets have made huge progresses since the first discoveries in the late nineties. In particular, the independent measurement of the mass and radius of planets, by combining the transit and radial-velociy techniques, allowed exploring their density and hence, their internal structure. With CoRoT (2007-2012), the pioneering CNES space-based mission in this investigation, about thirty new planets were characterized. CoRoT has enhanced the diversity of giant exoplanets and discovered the first telluric exoplanet. Following CoRoT, the NASA Kepler mission has extended our knowledge to small-size planets, multiple systems and planets orbiting binaries. Exploring these new worlds will continue with the NASA/TESS (2017) and ESA/PLATO (2024) missions. \\

\end{abstract}

\end{frontmatter}

\section{Introduction}

The search for exoplanets around solar-type stars has started in the 1990's with the direct imaging method and the radial velocity method. While the first one uses high-contrast and high-resolution imaging of nearby stars to search for the dim light of a sub-stellar companion, the second method uses indirect measurements of stars and searches for velocity wobbling due to the gravitational pull by an invisible planet. Both methods had their first results in 1995: the first detected brown dwarf companion Gl 229 B (about 50 Jupiter masses) by direct imaging \cite{oppenheimer} and the first detected Jupiter-like exoplanet  51 Pegasi b, orbiting a solar-type star \cite{mayor95}. The main difference between these two companions is probably their distance to their parent star: while the orbital period of Gl 229 B is larger than 10,000 years, the one of 51 Pegasi b is 4.5 days. This exemplifies the different biases of the two methods, imaging being sensitive to planets at long orbital period whereas radial velocity is performant for systems at short orbital period. In the meantime, the Hubble Space Telescope was discovering that young protostars were being formed in opaque, dusty disk structures, another hint that planetary formation could be a universal phenomenon.

These discoveries triggered a new field of research and extreme enthusiasm in the astrophysical community. A third method is quickly proposed and experimented on the first detected radial-velocity planets: it consists in searching for a slight dimming of the star's light that would be due to the crossing of a planet on the stellar disk -- a transit. This method works for planets whose orbit is perfectly aligned with the line of sight between the observer and the parent star. To discover transiting planets, it is thus necessary to observe for a long time a very large number of stars. The method is, as the radial-velocity method, biased towards detecting short-period planets. 

A few months after the discovery of 51 Peg b, a team of French astronomers and CNES engineers  proposed the spaced-based CoRoT mission \cite{baglin2003}, aimed at observing large numbers of stars with extreme photometric precision and long time series -as only possible from space-, in search for exoplanet transits. While CoRoT was designed, built, and integrated, a few exoplanetary transits were being detected from ground-based observatories. The first of them is HD 209458 b \cite{charbonneau}, a short-period Jupiter-like planet as 51 Pegasi b. The independent detections of HD 209458 b with the radial velocity and the transit methods have been a strong confirmation that these giant planets in very close distance to their stars actually existed. The combination of both methods also allowed to measure the planet's mass (from the radial-velocity amplitude) and its radius (from the transit depth), hence the bulk density of the planet. While the density of HD 209458 b is within the range of the Solar System's largest gaseous planets, the planet radius could not be reproduced by models of internal structure: the giant exoplanet appeared largely inflated, for its star's composition and age. 
When CoRoT was then launched in 2006, there were 15 known transiting exoplanets; all were giant, gaseous planets, whose transits (1-3\% deep, 2-3 hour long) were easily detected from ground-based telescopes of small size (typically less than 1m). 

In this article, we will review the findings of the CoRoT mission in its search for exoplanets. After describing the mission concept and instrument, we will review the process that starts with candidates and leads to exoplanets. Finally, we will emphasize the learnings from the CoRoT exoplanets and have a look towards other exoplanet surveys and future missions in this field.

\section{The CoRoT satellite}
Searching for exoplanets with the transit method requires both very good photometric precision and a very large number of observed stars. The  duration of the observing run sets the domain of orbital periods that is within reach, as the transit occurs once per orbital period (note that the secondary transit, occurring when the planet passes behind the star with respect to the observer, is only rarely detected due to a much smaller amplitude and additional constraint on the geometry). In addition to transit detection, precise stellar photometry from space also allows to probe the stellar interiors with asteroseismology, another main science objective of CoRoT (on a smaller number of much brighter stars) \cite{baglin2003}. In the domain of transiting exoplanets, a space-based mission allows to detect transiting planets of smaller size and/or of longer orbital period than ground-based photometric surveys.\\

The CoRoT design can be summarized by \cite{auvergne2009}:
\begin{itemize}
\item a PROTEUS platform ensuring  a polar orbit at 900km altitude and 6-month continuous access to a given field of view
\item a 27-cm mirror in an off-axis telescope
\item a focal plane divided in two, with an asteroseismology channel and an exoplanet channel; a field of view of 1.4$^\circ$x2.8$^\circ$ for the exoplanet channel  and 2.3 arcsec pixel size 
\item an operation plan adapted to the satellite orbit, with a succession of short runs (15-30 days) and long runs (80 or 150 days)
\item two main fields of observation, located in two opposite regions in the Milky Way towards the Monoceros and Aquila constellations and close to the Galactic plane. These locations were optimized for stellar density, considering other observing constraints, such as the presence of bright nearby stars of scientific interest for the asteroseismology channel
\item a 7.10$^{-4}$ photometric precision on a R=15 star in a 2h timescale
\item about 6,000 stars observed simultaneously in the exoplanet channel, in the magnitude range of V=11 to 16.5
\item 32-sec and 8-min temporal sampling on the light curves calculated on-board, and a duty cycle of 91\% 
\item for the 30\% brightest stars, three-color light curves generated by a low-resolution prism. Only the white light curve is provided for all other stars
\end{itemize}

Launched on December 26, 2006 by a Soyouz rocket from Baikonour, the CoRoT satellite has provided astronomical data from February 2007 (after a run of commissioning) to November 2012. Its lifetime was originally granted to be 3 years, and was  extended twice. Observations were discontinued due to electronic failures, probably due to high-energy particle bombardment. The satellite was stopped in June 2014 after programming its slow decay to Earth. 

A picture of the satellite during its integration in Cannes is shown on Figure 1. An archive of the CoRoT data is now fully public and can be accessed at http://idoc-corotn2-public.ias.u-psud.fr.

\begin{figure}
\includegraphics[width=\linewidth]{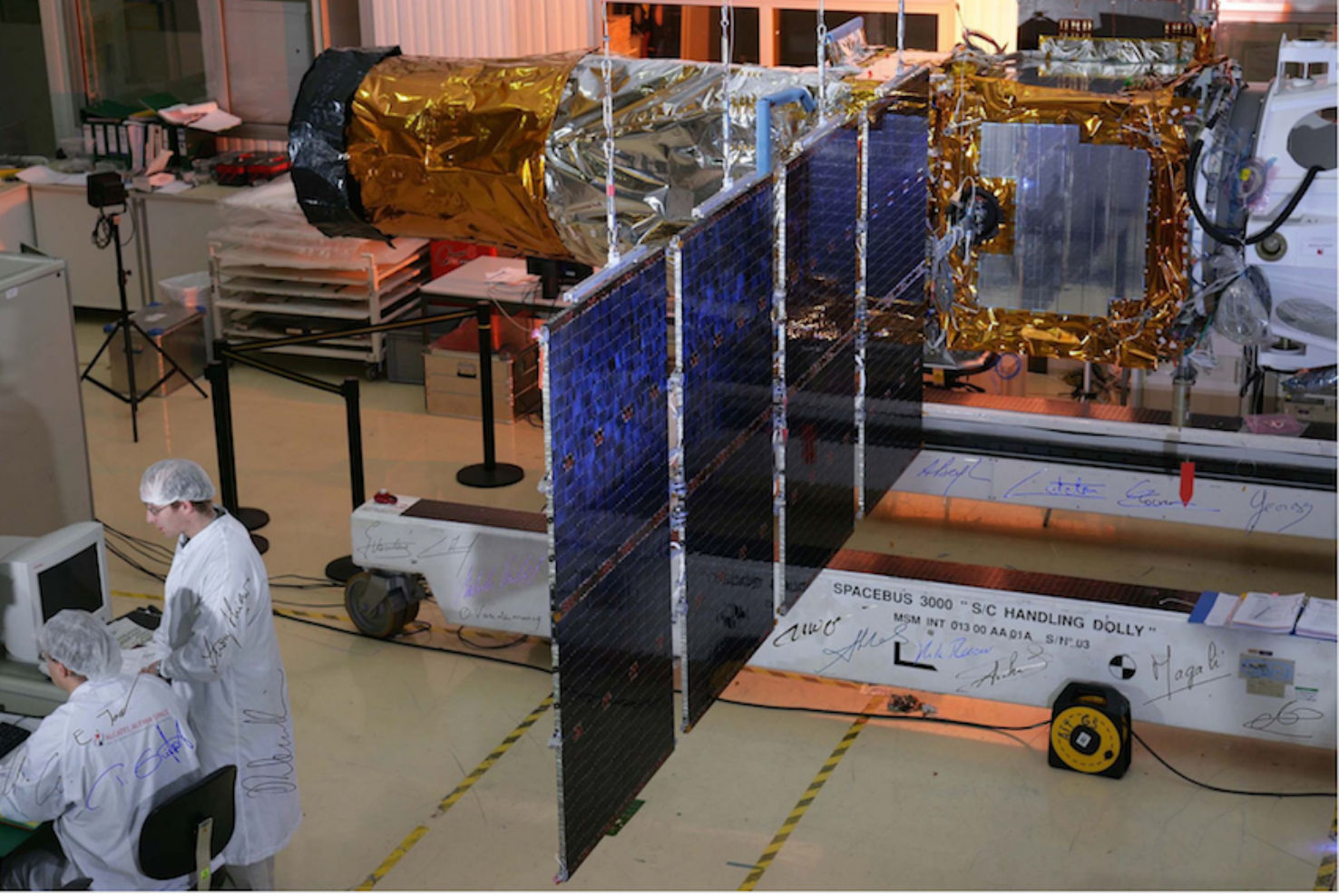}
\caption{The CoRoT satellite in its integration hall at Thales Alenia Space in Cannes, in 2005. 
}
\end{figure}

\section{Candidates or planets?}

In total, 163,664 stars were observed during the lifetime of CoRoT for a total of 169,967 light curves. Light curves were processed onboard and sent to Earth where additional data reduction got rid of instrumental effects. Several algorithms for transit detection were then applied to the light curves, generating several lists of exoplanet candidates. 

A planet candidate is not necessary a planet. There are several other astrophysical scenarios than mimic the planetary transits, involving multiple eclipsing stars rather than planets: grazing binaries, giant-dwarf binaries, eclipsing binary diluted by a third body. If the detected transit is very shallow, a possible contamination is a giant transiting planet orbiting a star diluted by another star. In such a case, the planet interpretation is correct, but the inferred radius of the planet would be largely incorrect.

As CoRoT provides lists of transiting candidates, these are then submitted to series of tests to evaluate their origin. For instance, the three-colored light curves, when available, can be used to eliminate candidates showing different transit depth in the different colors (signature of a blended stellar scenario). The distance between transits in a light curve giving the orbital period, it is possible to evaluate the expected transit duration from this period and the stellar radius; if too discrepant, the chances are high that the transit is due to a diluted eclipsing binary with a giant primary star. For each run, about 50 candidates are detected, and about half of them are further considered as compatible with a planetary signature. On these ones, it is then required to perform additional observations to unveil their nature and complete their characterization. Table 1 gives a summary of the number of CoRoT observing runs, observed stars, detected candidates and final planet yield in both Galactic directions. 

\begin{table}
\begin{center}
\caption{Stars, observing runs, light curves, candidates and planets counts from the CoRoT Exoplanet program. 
}
\begin{tabular}{lllll}
\hline
Direction      & Runs &  Light curves & Candidates & Planets \\
Aquila          &           13                &        85489       &         348       & 17 \\
Monoceros &           13             &            89215      &         251       &   16 \\ 
\hline
\end{tabular}
\end{center}
\end{table}

Complementary observations of CoRoT candidates include time-critical ground-based photometry, adaptive optics, and high-resolution spectroscopy with a high radial-velocity precision instrument. The first observation provides us with a confirmation that the transit does not occur on a nearby star (less than 30 arc sec away); it is required by the fact that CoRoT pixels and CoRoT photometric apertures are, respectively, 2.3 arcsec and $\sim$ 30 arcsec wide, and the observed star can be contaminated by fainter neighbors. Adaptive optics follows up on this objective, putting constraints on fainter neighbors at shorter distances. Finally, a radial velocity monitoring of the target identifies the grazing binaries, the multi-component spectroscopic binaries, and the planets with a measurable mass. The magnitude range of CoRoT targets, defined to have an appropriate stellar sample size due to geometrical restrictions of the transit alignment, severely restricts the capability of  spectroscopic follow-up observations, especially for stars fainter than 14. This is even more critical even for very shallow transits that are potentially due to very low-mass planets. Nevertheless, more than 120 nights of telescope time with high-precision radial-velocity spectrographs were dedicated to complementary observations of CoRoT candidates, including HARPS at the European Southern Observatory, SOPHIE at Observatoire de Haute Provence, HIRES at Keck Observatory and FIES on the Nordic Optical Telescope. The outputs of complementary observations are among the following:

\begin{itemize}
\item another neighbor star is identified in the vicinity of the main target, which shows deep, stellar eclipses at the expected time. The  transit detected by CoRoT is the result of a multiple-system dilution and usually not considered further.
\item the spectrum shows two or more components, or an amplitude of the radial-velocity curve consistent with a companion in the stellar mass regime and at the correct period. This  candidate planet is rejected.
\item the spectrum shows a single component, and the radial-velocity curve reveals a mass of the transiting companion which is compatible with a planetary nature. Then a full characterization of the planetary system is undertaken.
\item the spectrum reveals a fast rotating star, or a star hotter than $\sim$8000K, for which radial velocity precision is too limited for a planet characterization. The status of the candidate remains undefined.
\item the radial-velocity series has sufficient precision but no signature is revealed at the ephemeris of the CoRoT transit signals. Unless a complex, peculiar system is identified, the system's nature remains unclear and an upper limit for the mass of a possible planetary candidate is estimated. 
\end{itemize}

\begin{figure}
\includegraphics[width=\linewidth]{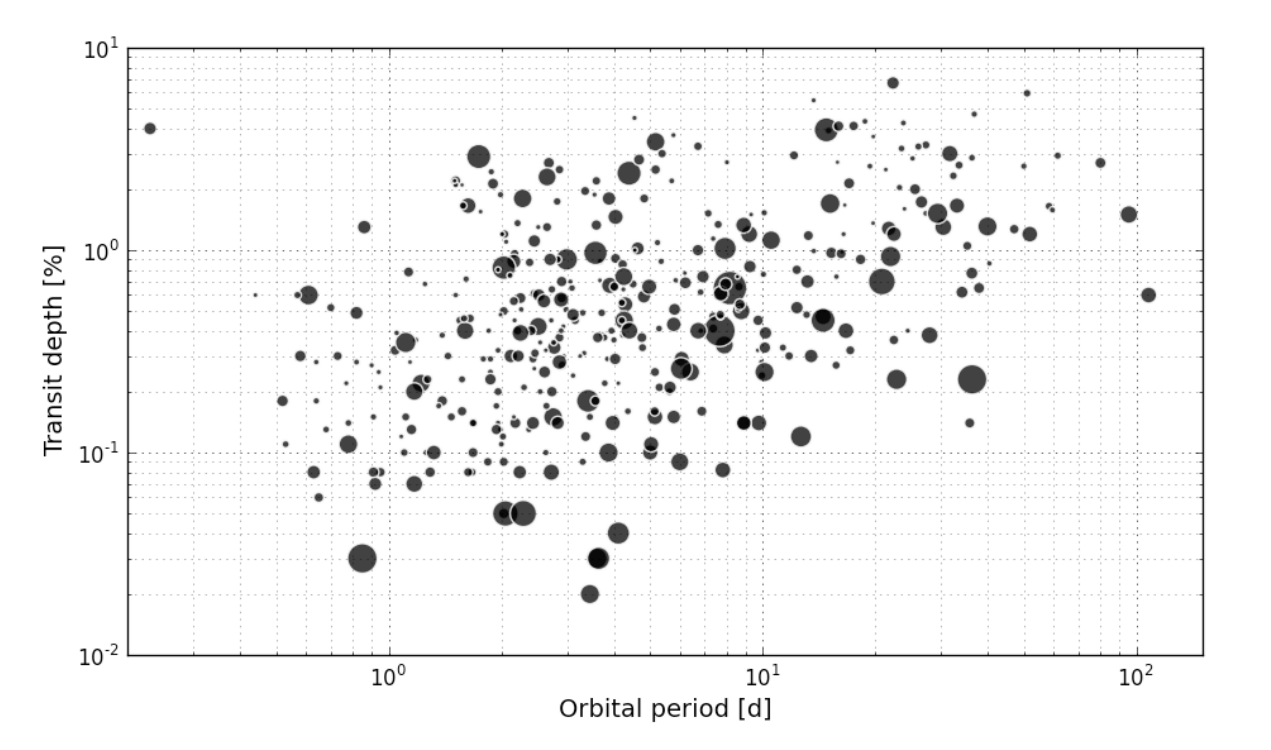}
\caption{The transit depth versus the orbital period of CoRoT candidates. The symbol size is  proportional to the star brightness. 
(illustration by A. Santerne)}
\end{figure}

In the remaining cases where complementary observations bring no definite answer, we have to rely on statistical comparison of data with a set of different astrophysical models. The CoRoT team has developed such a tool using Bayesian statistics, PASTIS \cite{diaz2014}, which has played a large role in some planet identifications, such as CoRoT-16 b, CoRoT-22 b and CoRoT-24 b.

\section{CoRoT main discoveries}
As for other subsamples of exoplanets, the CoRoT discoveries show an extreme diversity of parameters, especially when compared to planets in the Solar System. First, as CoRoT measures the planetary radii, it is directly observed that some planets have large radius anomalies. These anomalies can go in both directions: either positive (radius inflation) or negative (radius contraction) and are related to the theories of formation and evolution of a solar-composition gaseous planet made of hydrogen and helium, with no core. Several planets characterized by CoRoT and its follow-up instruments have a radius larger by more than 10\%: CoRoT-1 b, 2 b, 5 b, 11 b, 12 b, 19 b \cite{moutou2013}, and 26 b. This inflation is thought to be due to a small fraction of additional energy received by the planet and dissipated in its interior \cite{guillot2002}. Several mechanisms have been proposed for these non-radiative processes, for instance invoking tidal interactions with the close-by star. If most exoplanet inflated radii can be explained with a very small amount of additional energy (about 1\%), one of them, CoRoT-2 b, requires much more because of its large mass. Other mechanisms have been proposed, invoking dynamical events during the past evolution of  this system \cite{guillot2011}. 

Other planets have a negative radius anomaly; their radius is smaller than expected with respect to their mass, expected composition and age. This anomaly sometimes requires huge amounts of heavy elements, probably condensed in a central core. The dense planets discovered by CoRoT are: CoRoT-8 b, 10 b, 13 b, 20 b. CoRoT-20 b \cite{deleuil2012} is an extreme case, requiring about 500 Earth masses in heavy elements and questioning the formation mechanism of planets with such heavy elements  enrichment.

CoRoT-9 b \cite{deeg2009} has attracted interest, among CoRoT discoveries, because of its relatively long orbital period (95 days). It is a giant planet and its radius is not inflated with respect to the expect H-He composition, illustrating the fact that radius inflation in other transiting giant planets was indeed due to excess energy income from their proximity to the star.  

CoRoT also discovered the first transiting planet in the super-Earth regime, CoRoT-7 b \cite{leger2009}. The planet is qualified as  Super-Earth because it has a density of 6.61 g/cm$^3$, close to the density of Earth \cite{haywood2014}, and a radius of $\sim$10,000km (1.58 Earth radius) \cite{barros2014}. The discovery of CoRoT-7 b triggered a large number of complementary observations and studies. Because it is orbiting an active star, the radial-velocity observations are contaminated by stellar intrinsic variations modulated by the rotation period of the parent star; in addition, another non-transiting planet at 3.7 day period contributes to the stellar wobble. These signals complicate deriving the mass of the inner telluric planet \cite{haywood2014}. CoRoT-7 b is also a challenge to planetary science by its extremely short period: 0.85 day. Its distance to the surface of its star is only 3.4 times the stellar radius, which implies extreme irradiation fluxes and strong tidal interactions. The planet spin and orbit being likely synchronized, the day side surface may be a deep lava ocean, while the night side would be frozen, with a temperature difference between both hemispheres reaching 2400K \cite{leger2011}. The composition of CoRoT-7 b is likely dominated by silicates, that are eroded into a thin atmosphere of rocky vapor due to the large surface temperatures \cite{leger2011}. Today, thanks to Kepler and ground-based surveys around M stars, a small number of similar planets have been discovered.

CoRoT has found another multiple planetary system composed of two transiting planets or 3.7 and 5 Earth radii and 5.1 and 11.8 day periods, respectively \cite{alonso2014}. Both are compatible with Neptune-like planets, with a significant gaseous envelope. There may be a third, massive planet in the same system, at an orbital distance at least 30 times larger. Except for CoRoT-7 and CoRoT-24 systems, all other planets discovered by CoRoT seem to be single companions of their parent star. CoRoT has found no planet orbiting binary stars.

\begin{figure}
\includegraphics[width=\linewidth]{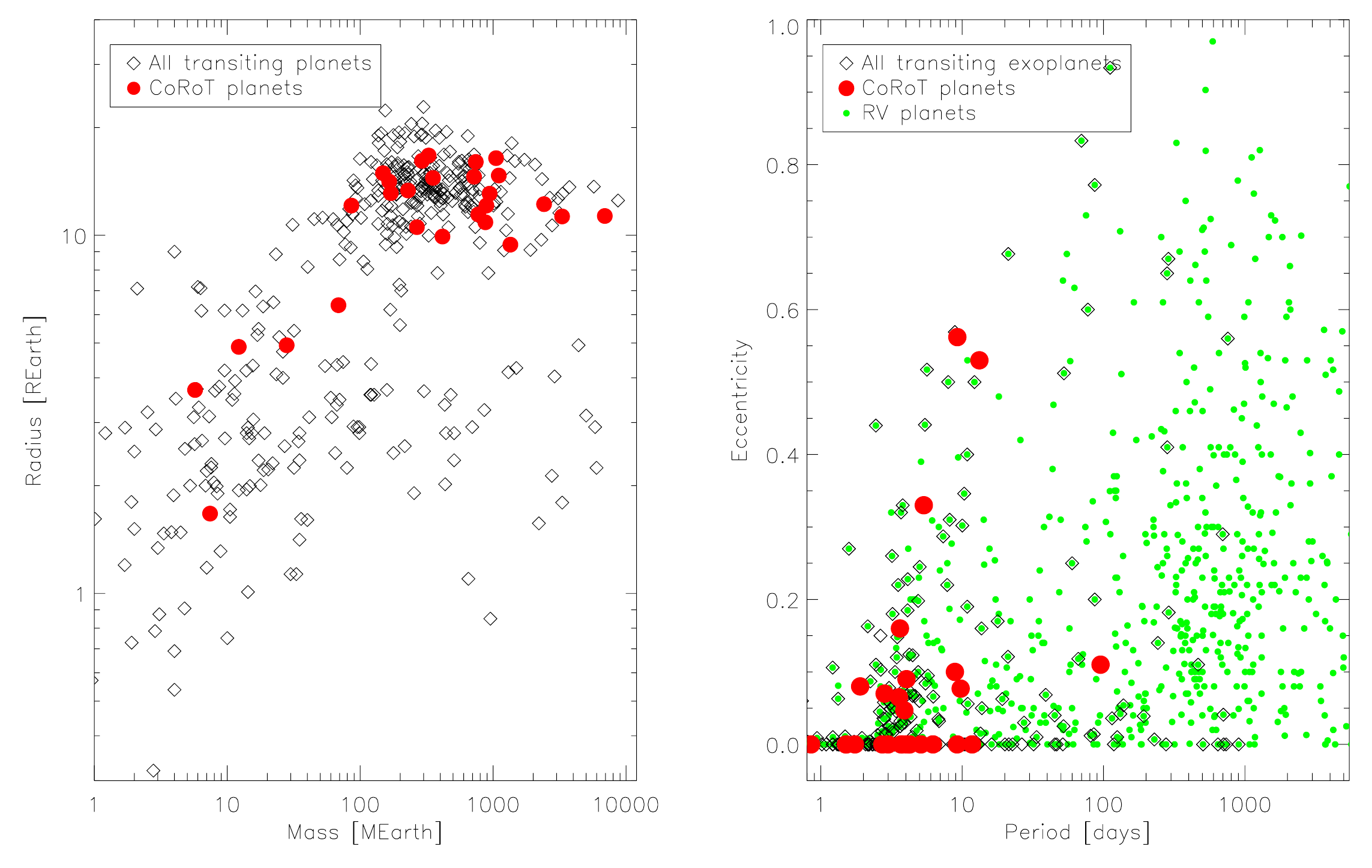}
\caption{The mass and radius of the CoRoT planets (left) and the orbital eccentricity as a function of period (right). Open diamonds show all transiting planets known at the beginning of 2015, red filled circles show CoRoT planets, and small green dots show non-transiting radial-velocity planets.
}
\label{mr}
\end{figure}

As a summary, the mass-radius and the period-eccentricity diagrams of CoRoT exoplanets are shown on Figure \ref{mr}. Superimposed to CoRoT exoplanet parameters are exoplanet properties measured with other surveys until January 2015. The diversity observed with CoRoT has been confirmed by other observing campaigns, and even further extended.

\section{Future is bright}
CoRoT is not the only transit-search survey in this past decade. Simultaneously to the preparation and launch of CoRoT, several ground-based surveys have started, to search for exoplanets. These surveys (eg, Super-WASP, HAT, TrES, Qatar), if biased towards giant exoplanets with short periods, were extremely productive and largely contributed to our current knowledge of these planetary companions (Figure \ref{mr}).

In March 2009, NASA launched the Kepler satellite, a more ambitious, large-scale mission for exoplanet transit searches. The outcome of the Kepler mission has shown the ubiquity of small-size planets and the existence of compact aligned multiple systems \cite{batalha,lissauer,marcy}. Kepler stopped observations in May 2013 and the telescope resumed observations in a lower-performance mode, under the name K2, in June 2014.

The limitations of the pioneer space transit surveys undertaken by CoRoT and Kepler are related to the magnitude range of the target stars. For both missions, the transit candidates orbiting the most numerous faint stars could not be followed up with the highest accuracy radial velocity spectrographs, and only an upper limit of their mass is (or will be) available. This limitation will not exist anymore for the next space missions TESS (NASA, launch in 2017) and PLATO (ESA, launch in 2024). Focusing on the brightest stars in a much larger part of the sky, TESS and PLATO will have the capacity of unveiling the low-mass planets in the habitable zones of their stars (cool stars for TESS, solar-type stars for PLATO), and to characterize these stars totally through intense complementary observing campaigns. Getting the planet masses with a few \% accuracy in the domain of Earth masses is challenging, but possible if the parent star is bright enough. New facilities will be added in the next years to the existing high-precision radial-velocity instruments (VLT/ESPRESSO, CFHT/SPIRou, Calar-Alto/CARMENES, Subaru/IRD...), and their contribution to TESS and PLATO complementary observations will be highly required. The field of research pioneered by CoRoT has thus an  inestimable future. The next step after identifying and characterizing  other terrestrial worlds in the solar neighbourhood will be to explore their atmospheric properties, in particular in a quest for biosignatures.\\

{\bf Acknowledgements}\\
This is an invited contribution to the special issue "Invited contributions of 2014 geoscience laureates of the French Academy of Sciences ". It has been reviewed by Fran\c{c}oise Combes and editor Vincent Courtillot. We are thankful to the technical teams that have been in charge of CoRoT at CNES and in partner institutes from design to operations. We want to acknowledge the role of the science team, for its high motivation for about ten years. CoRoT would not have been such an adventure without the far-reaching involvement of A. Baglin, J. Schneider, A. L\'eger and D. Rouan particularly. \\

\bibliography{mybibfile}

\end{document}